\documentclass[twocolumn,showpacs,preprintnumbers,amsmath,amssymb,pra]{revtex4}
\usepackage{graphicx}% Include figure files
\usepackage{dcolumn}% Align table columns on decimal point
\usepackage{bm}% bold math
\usepackage{epsfig}

\newcommand{\beq}{\begin{equation}}
\newcommand{\eeq}{\end{equation}}
\newcommand{\bea}{\begin{array}}
\newcommand{\ena}{\end{array}}
\newcommand{\beqa}{\begin{eqnarray}}
\newcommand{\eeqa}{\end{eqnarray}}

\newcommand{\qq}{\mathbf{q}}
\newcommand{\pp}{\mathbf{p}}

\begin{document}
\title{Bose-Einstein condensation temperature of a gas of weakly dissociated diatomic molecules}
\author{L. M. Jensen}
\affiliation{Department of Physics, Nanoscience Center, P. O. Box 35, FIN-40014 University of Jyv\"askyl\"a, Finland}
\author{H. M\"akel\"a}
\affiliation{Department of Physics, University of Turku, FI-20014 Turku, Finland}
\author{C. J. Pethick}
\affiliation{Nordita, Blegdamsvej 17, DK-2100 Copenhagen \O, Denmark}

\begin{abstract}
We consider the properties of a gas of bosonic diatomic molecules in
the limit when few of the molecules are dissociated.  Taking into
account the effects of dissociation and scattering among molecules and
atoms, we calculate the dispersion relation for a molecule, and the
thermal depletion of the condensate.  We calculate the dependence of
the Bose-Einstein condensation temperature of a uniform gas on the
atom-atom scattering length, and conclude that, for a broad Feshbach
resonance, the condensation temperature increases as the molecular
state becomes less strongly bound, thereby giving rise to a maximum in
the transition temperature in the BEC-BCS crossover. We also argue on
general grounds that, for a gas in a harmonic trap and for a narrow
Feshbach resonance, the condensation temperature will decrease with
increasing scattering length.

\end{abstract}
\maketitle                       

\section{Introduction}

The crossover between BCS pairing of two species of fermion and
Bose-Einstein condensation of a diatomic molecules with changing
atom-atom scattering length has been studied by many authors,
beginning with the works of Eagles \cite{eagles}, Leggett
\cite{leggett} and Nozi\`eres and Schmitt-Rink \cite{nozieres}.  One
of the surprising results of the calculations of Ref.  \cite{nozieres}
and also found in Refs. \cite{Drechsler1992,Melo1993} is that, in the
crossover region, the transition temperature has a maximum as the
strength of the atom-atom interaction is varied.  However, it is
unclear whether the maximum is real, or merely an unphysical artifact
of either the assumptions made in the model or the numerical
procedures.  The early calculations only accounted for quadratic
fluctuations around the mean-field solution and it was generally
expected that the fluctuation induced peak was an artifact of the
gaussian approximation. This view was furthermore supported by a
self-consistent field theoretical calculation in Ref.
\cite{Haussmann1994} where $T_{\rm c}$ was found to be a monotonic function
of the dimensionless coupling constant. However, the most recent
self-consistent calculations of $T_{\rm c}$  \cite{Haussmann2006} do indicate
the presence of a peak in $T_{\rm c}$.  Most calculations for broad
resonances for a uniform gas give a peak in $T_{\rm c}$ for both one-channel
and two-channel models for the interaction, and for a variety of
approximations for the atom propagators, which are in some cases
replaced by either bare propagators, or by dressed propagators
\cite{Milstein2002,Ohashi2002,Ohashi2003}. In
calculations that use a combination of bare and dressed propagators,
$T_{\rm c}$ initially decreases coming from the far BEC limit before turning
around to give a local maximum near resonance \cite{levin}.  In calculations
for a harmonically trapped gas in the local density approximation, the
peak vanishes for both one-channel and two-channel models for the
interaction, and for a variety of approximations for the atom
propagators \cite{Perali2004}.  The peak vanishes in calculations for
sufficiently narrow resonances.  A summary of the present situation is that
microscopic calculations point to there being a maximum in $T_{\rm c}$ for broad resonances for the uniform gas,
but not for harmonically trapped gases or sufficiently narrow resonances. 

The purpose of this paper is to study the properties of a weakly
dissociated gas of diatomic molecules composed of fermionic atoms with
a view to determining whether or not the transition temperature
$T_{\rm c}$ to a paired state should exhibit a maximum in the BEC-BCS
crossover.  We shall consider the molecule effective mass and the
reduction of the density of molecules due to dissociation, as well as
the interaction between molecules.  As a consequence of critical
fluctuations, molecule-molecule interactions lead to an increase in
$T_{\rm c}$ away from the BEC limit and this effect dominates the
decreases in $T_{\rm c}$ we find due to changes in the effective mass
and the number of molecules.
For a trapped gas, we shall argue that $T_{\rm c}$ should decrease away from 
the BEC limit due to the reduction of the central density in the trap as a consequence 
of the increasingly repulsive molecule-molecule interaction.

This paper is organized as follows.  In Sec. II we consider the
uniform system, and calculate the effective mass of a molecule and
thermal depletion of the number of molecules.  The transition
temperature is discussed in Sec. III, first for a uniform gas and then
for a trapped gas.  Finally, Sec. IV contains concluding remarks.

\section{The uniform system}
 
We investigate the properties of a system containing equal
numbers $N$ of two species of fermion, labelled by a pseudospin index
$\sigma=\pm 1$, that can form a diatomic molecular
bound state in the limit when the system consists mainly of molecules,
with a small admixture of atoms. We shall assume that the two species
of fermion have the same mass $m$. In the spirit of
Landau, we write the effective low-energy Hamiltonian for the system as
\begin{eqnarray}
 H &=& \sum_{\pp}\left(-\epsilon_\mathrm{b}+\frac{p^2}{4m}\right)\hat{b}^\dag_{\pp}
\hat{b}_{\pp} +
\sum_{\qq \sigma}\frac{q^2}{2m}\hat{a}^\dag_{\qq\sigma}
\hat{a}_{\qq\sigma}\nonumber \\
 &+ &   \frac{\pi\hbar^2a_{\rm mm}}{mV}\sum_{\pp\pp'\qq}
\hat{b}^\dag_{\pp}\hat{b}^\dag_{\pp'}\hat{b}_{\pp'+\qq}
\hat{b}_{\pp-\qq}\nonumber\\
& + &\frac{6\pi\hbar^2a_{\rm am}}{mV}\sum_{\pp\pp' \qq \sigma}
\hat{a}^\dag_{\pp\sigma}\hat{b}^\dag_{\pp'}\hat{b}_{\pp'+\qq}
\hat{a}_{\pp-\qq\sigma}\nonumber \\
&+&\frac{4\pi\hbar^2a_{\rm aa}}{mV}\sum_{\pp\pp'}
\hat{a}^\dag_{\pp\uparrow}\hat{a}^\dag_{\pp'\downarrow}\hat{a}_{\pp'+\qq\downarrow}
\hat{a}_{\pp-\qq\uparrow}\nonumber\\
 & + &\frac{g}{V^{1/2}}\sum_{\pp\qq}\left(\hat{b}^\dag_{\pp}
\hat{a}_{\qq+\pp/2\uparrow}\hat{a}_{-\qq+\pp/2\downarrow}+
\mathrm{h.c.}\right)
%\hat{b}_\pp \hat{a}^\dag_{\qq+\pp/2,\uparrow}\hat{a}^\dag_{-\qq+\pp/2,\downarrow}),
\end{eqnarray}
the operators $\hat a^\dag$, $\hat a$     $\hat b^\dag$, $\hat b$ create and destroy atoms and molecules, respectively.   Here $V$ is the volume of the system, $g$ is the matrix element for
decay of a molecule into
two atoms and $a_{\rm ij}$ is the scattering length between a particle
i and a particle j, and `a' stands for an atom and `m' for a molecule.
The  quantity $a_{\rm aa}$ is the background scattering length for atom-atom scattering
due to nonresonant processes.
In the case of strong coupling between the molecule and the atoms,
e.\ g.\ for a broad Feshbach resonance, and in the limit of a
deeply bound molecular state it has been shown that
\cite{termartirosian,petrov}
\begin{eqnarray}
 a_{\rm ma} \approx 1.2a_{\rm aa}, &\,\,\, & a_{\rm mm}\approx 0.6a_{\rm aa}.
\end{eqnarray}
%and \cite{petrov}
%\beq
% a_{\rm mm}\approx 0.6a_{\rm aa}.
%\eeq

We begin by considering a uniform gas, and we first calculate the
molecule spectrum.    The leading contributions to the energy of a
molecule are due to the molecule-molecule interactions, which give a
shift $2\pi \hbar^2 n_{\rm m}a_{\rm mm}/m$ to the self energy of a
molecule in the mean field of the other molecules, and a contribution
one half of this to the average energy per molecule. Here $n_{\rm m}$ is
the number density of molecules.  Atom-molecule
interactions contribute $6\pi \hbar^2 n_{\rm a}a_{\rm am}/m$ to the
molecule self energy from the scattering of atoms by molecules, and
from dissociation of molecules into atoms an amount
\beqa
\delta \epsilon_{\pp}=g^2\int \frac{d^3q}{(2\pi\hbar)^3}
\frac{n_{\pp/2+\qq} +n_{\pp/2-\qq}}{\epsilon_\mathrm{b}+q^2/m
}\nonumber \\
=2g^2\int \frac{d^3q}{(2\pi\hbar)^3}
\frac{n_\qq}{\epsilon_\mathrm{b}+(\qq+\pp/2)^2/m}.
\label{energyshift}
\eeqa
The term independent of the atom distribution function $n_{\qq}$ is not
included, since this is already contained in the binding energy $\epsilon_\mathrm{b}$ 
of two fermions in the absence of a medium. The contribution (\ref{energyshift}) is due to Pauli
blocking of dissociation of molecules, which raises
the energy of a molecule.  As a consequence of Galilean invariance, in
the absence of a medium, the effective
mass of a molecule remains equal to the bare mass.
The binding energy of a molecule with momentum zero is thus reduced by an amount
\beq
\delta \epsilon_\mathrm{b} = 2g^2\int \frac{d^3q}{(2\pi\hbar)^3}
\frac{n_\qq}{\epsilon_\mathrm{b} +q^2/m}\simeq   2g^2
\frac{n_{\rm a}}{\epsilon_\mathrm{b}},
\eeq
where $n_{\rm a}$ is the density of atoms of both species.
In the second expression we have assumed that the molecule is
deeply bound in the sense that $\epsilon_\mathrm{b}\gg kT$.  The molecule
spectrum varies over momentum scales of order $(m\epsilon_\mathrm{b})^{1/2}
$, and therefore for the momenta of interest in the problem of
Bose-Einstein condensation, which are of order  $(mkT)^{1/2}$, it is
sufficient to calculate only to order $p^2$.  The result is
\beq
\epsilon_\pp\simeq - \epsilon_\mathrm{b}  +  \frac{2\pi \hbar^2 n_{\rm m}a_{\rm
mm}}{m}+ \frac{6\pi \hbar^2 n_{\rm a}a_{\rm am}}{m}  + 2g^2
\frac{n_{\rm a}}{\epsilon_\mathrm{b}}+
\frac{p^2}{2M^*},
\eeq
where the effective mass of a molecule is increased by an amount
given by
\beq
\frac{1}{M^*}=\frac{1}{2m} \left(1-2g^2
\frac{n_{\rm a}}{\epsilon_\mathrm{b}^2}\right).
\eeq
We now turn to the transition temperature of the gas.

\section{Transition temperature}

\subsection{Uniform system}
For a uniform gas of elementary bosons, the transition temperature is
unaffected by the Hartree mean-field energy, since this gives a shift
to the energy of a boson and to the chemical potential which are
independent of the momentum of the particle.  However,
it has been demonstrated that, due to critical fluctuations, the
transition temperature is increased, by an amount which is given for
small $n_{\rm m}^{1/3}a_{\rm mm}$  by \cite{baym,footnote1}
\beq
    \frac{\delta T_{\rm c}}{T_{\rm c}}=b  n_{\rm m}^{1/3}a_{\rm mm},
\label{Tcfluct}
\eeq
where $b$ is a positive coefficient, which has been estimated to be
approximately $1.3$ \cite{arnold, kashurnikov, footnote2}.  This effect will tend
to be counteracted by
the increase of the effective mass and the reduction of the density of
molecules due to dissociation, both of which tend to decrease the
transition temperature, since for an ideal Bose gas, $T_{\rm c}\sim
\hbar^2n_{\rm m}^{2/3}/2M^*$.  The number density of molecules is given
by
\beq
n_{\rm m}=\frac{N}{V}-\frac{n_{\rm a}}{2}.
\eeq
The two latter effects give changes in the transition temperature which
are proportional to $n_{\rm a}/n_{\rm m}\propto e^{-\epsilon_\mathrm{b}/kT_{\rm
c}}$.  Thus for a deeply bound molecule, the effects of critical
fluctuations dominate, and the transition temperature initially increases away
from the BEC limit.  Since in the BCS
limit the transition temperature tends to zero, we conclude that the
transition temperature must have a local maximum, provided that the
transition temperature is a continuous function of the scattering
length.

\subsection{Trapped cloud}

We now turn to the case of a cloud in a harmonic trap, and we shall
assume that the trapping potentials for the two species of fermion are
the same, and equal to one half of the trapping potential for a
molecule. Previously, the transition temperature for a
trapped gas of bosons has been calculated with allowance for the nonzero
spacing of the single-particle levels in the oscillator potential and
for the distortion of the density profile by the interaction and the
result is \cite{Giorgini1996}
\beq
    \frac{\delta T_{\rm c}}{T_{\rm c}}\approx -0.73
\frac{\omega_{\rm m}}{\bar \omega} N^{-1/3}-1.33\frac{a_{\rm mm}}{\bar
a}N^{1/6},
\label{Tcnofluct}
\eeq
where $\omega_{\rm m}$ is the algebraic mean of the trap frequencies for
the three principal axes of the trap,  $\bar \omega$ is their
harmonic mean and $\bar a = (\hbar/m\bar \omega)^{1/2}$ is the mean harmonic oscillator length.  

To take account of critical fluctuations, we argue that provided the
typical  interaction energy between molecules, $2\pi \hbar^2 n_{\rm
m} a_{\mathrm{mm}}/m$
is large compared with the oscillator quantum of energy but small
compared with $kT_{\rm c}$, it should be a good approximation to
calculate the change in $T_{\rm c}$ in a local density approximation,
in which the central density of the cloud is inserted in Eq.\
(\ref{Tcfluct}).  The central density  $ n_{\rm m}(0)$     at $T_{\rm c}$ is given by
\beq
  n_{\rm m}(0) =
\zeta(3/2)\left(\frac{MkT_{\rm c}}{2\pi\hbar^2}\right)^{3/2},
\eeq
and since $kT_{\rm c}=[N/\zeta(3)]^{1/3}\hbar{\bar \omega}$,
it follows that the contribution to the change in $T_{\rm c}$ due to
critical fluctuations is
\beq
   \frac{\delta T_{\rm c}}{T_{\rm c}}\approx \frac{1.3}{(2\pi)^{1/2}}
\frac{[\zeta(3/2)]^{1/3}}{[\zeta(3)]^{1/6}}N^{1/6}\frac{a_\mathrm{mm}}{\bar a}.
\eeq
Calculations indicate that critical fluctuations are suppressed in inhomogeneous systems \cite{zobay2005},  
so we regard this   estimate as an upper limit to the increase of $T_{\rm c}$.
Adding this to the earlier result (\ref{Tcnofluct}) without
fluctuations, we find for the total change in $T_{\rm c}$
\beq
          \frac{\delta T_{\rm c}}{T_{\rm c}}\approx -0.73
\frac{\omega_{\rm m}}{\bar \omega} N^{-1/3}-0.64\frac{a_{\rm mm}}{\bar
a}N^{1/6}.
\eeq
Thus, in contrast to the uniform case, the change in $T_{\rm
c}$ for small
$a_{\rm mm}$ is
{\it negative}, since the effect of critical fluctuations is overwhelmed
by the reduction of the central density.

The discussion we have given above applies to broad Feshbach
resonances, for which the energy dependence of the atomic interactions
may be neglected.  For narrow Feshbach resonances, critical
fluctuations are suppressed because the magnitude of the
contribution from processes in which the argument of the interaction
contains a frequency which is to be summed over will be reduced
because of the strong reduction of the interaction away from
resonance, as discussed in Ref.\ \cite{schwenk}.  

\section{Conclusion}

In this paper we have identified a number of effects that produce
changes in the transition temperature away from the BEC limit.  Two of
these are due to thermal dissociation of molecules, and these lead to
a decrease in $T_{\rm c}$. However, for a uniform gas, these effects are
overwhelmed by a rise in $T_{\rm c}$ due to the increasingly repulsive
molecule-molecule interaction. We therefore conclude, quite generally,
that $T_{\rm c}$ must display a global maximum between the BEC and BCS
limits for a broad Feshbach resonance.  For trapped gases, we have argued that $T_{\rm c}$ will
initially decrease away from the BEC limit.  Thus, on the basis of the
arguments given here, there need not be a maximum in $T_{\rm c}$ but
it cannot be excluded.  Likewise, for a narrow Feshbach resonance, critical fluctuations are suppressed, and consequently
$T_{\rm c}$ is expected to decrease away from the BEC limit.

The framework we have described in this paper could be a useful tool
for analysis of detailed crossover calculations, since it points to a
number of quantities that could be calculated in order to test our
fundamental understanding.

%%\begin{acknowledgements}
This work was carried out under the auspices of a Nordita Nordic Project
on ultracold atoms, and H.\ M.\ wishes to thank Nordita for hospitality.
L. M. J. was supported by Nordita through a Nordic Fellowship and by the 
Academy of Finland and EUROHORCs (EURYI, Academy Projects, Nos. 106299, 205470, 
and 207083). H. M. was supported by the Finnish Academy of Science and 
Letters (Ville, Yrj\"o and Kalle V\"ais\"al\"a Foundation) and the Academy 
of Finland (grant no 206108). We are grateful to acknowledge G.\ Watanabe 
and  H.\ M.\ Nilsen for useful conversations.
%%\end{acknowledgements}


\begin{thebibliography}{99}
\bibitem{footnote2}In the calculations of the critical fluctuations,
the energy dependence of the effective interaction is neglected and consequently
the result is appropriate for a broad Feshbach resonance.  We shall
later return to the case of a narrow resonance.
\end{thebibliography}
\end{document}